\documentclass[11pt]{article}
\usepackage{amssymb,amsmath,epsf,graphicx}
\newcommand{\be}{\begin{equation}}
\newcommand{\ee}{\end{equation}}
\newcommand{\bea}{\begin{eqnarray}\displaystyle}
\newcommand{\eea}{\end{eqnarray}}
\newcommand{\bdm}{\begin{displaymath}}
\newcommand{\edm}{\end{displaymath}}
\newcommand{\sectiono}[1]{\section{#1}\setcounter{equation}{0}}

\newcommand{\Tr}{\mathop{\rm Tr}\nolimits}

\def\ket#1{|#1 \rangle}

\def\aver#1{\langle\, #1 \,\rangle}

\let\eps = \varepsilon

\def \ll {{\cal L}}

\def \bb {{\cal B}}

\def \sf  {string field }
\def \sft {string field theory }

\def \lll {{\widehat{\cal L}}}
\def \bbb {{\widehat{\cal B}}}
\def \Uhat {{\widehat{U}}}
\def \Lhat {{\widehat{\cal L}}}

\begin{document}
{}~ \hfill\vbox{\hbox{hep-th/0701248} }\break \vskip 2.5cm

\centerline{\Large \bf Comments on Marginal Deformations} \vspace*{2.0ex} \centerline{\Large \bf in
Open String Field Theory} \vspace*{8.0ex}

\centerline{\large \rm Martin Schnabl}

\vspace*{8.0ex}

\centerline{\large \it Institute for Advanced Study, Princeton, NJ 08540 USA} \vspace*{2.0ex}
\centerline{E-mail: {\tt schnabl at ias.edu}}

\vspace*{6.0ex}

\centerline{\bf Abstract}
\bigskip

In this short letter we present a class of remarkably simple solutions to Witten's open string
field theory that describe marginal deformations of the underlying boundary conformal field theory.
The solutions we consider correspond to dimension-one matter primary operators that have
non-singular operator products with themselves. We briefly discuss application to rolling tachyons.

 \vfill \eject

\baselineskip=16pt

%\tableofcontents
%\newpage

%%%%%%%%%%%%%%%%%%%%%%%%%%%%%%%%%%%%%%%%%%%%%%%%%%%%%%%%%%%%%%%%%%%%%%%%%%%%%%
\sectiono{Introduction}
\label{s_intro}
%%%%%%%%%%%%%%%%%%%%%%%%%%%%%%%%%%%%%%%%%%%%%%%%%%%%%%%%%%%%%%%%%%%%%%%%%%%%%%

One of the important features of string field theory is that it allows us to describe physics of
different string theory backgrounds using data of only a single reference conformal field theory.
This has been recently successfully applied in the context of Witten's open bosonic string field
theory \cite{Witten-SFT}. It has been shown, in accordance with Sen's conjectures \cite{SenConj},
that the theory formulated on an arbitrary D-brane describes another vacuum \cite{Bernoulli} with
no D-branes, and hence no conventional open string degrees of freedom \cite{ES}.\footnote{For
related recent development see \cite{Okawa,FK,RZspecial,ORZ,Erler1,Erler2}. Nice reviews of \sft
include \cite{TZ,Sen-review}.}

In this letter we shall give a construction of string field theory solutions that correspond to
less dramatic changes of the conformal field theory. Our exact solutions will describe conformal
field theories deformed by exactly marginal operators. We shall construct such solutions
perturbatively in a parameter $\lambda$, which to the first order can be identified with the
coupling constant of a given exactly marginal operator \cite{SenBI,SZBI}. Following
\cite{Bernoulli}, we shall use the cylinder conformal frame parameterized by a coordinate $\tilde z
= \arctan z$. The solution itself will be given by a series expansion in $\lambda$, each term will
be given by a cylinder with simple insertions of the $c$ ghost, the exactly marginal operator
called $J$, and vertical line insertions of the $b$ ghost. The mutual distances between $c J(z)$
insertion points will be parameters that will be integrated over. Unfortunately, for the generic
perturbation with singular operator product expansion our solution becomes ill-defined, so we shall
restrict ourselves mostly to cases in which $J(x)J(y)$ is finite when $x$ approaches $y$.

One of the more interesting examples of this kind is the time-dependent rolling tachyon solution
which is generated by exactly marginal operator $J(z)=e^{X^0(z)}$ studied in \cite{SenRolling}. We
will look at the time-dependent behavior of the tachyon coefficient to get some clues on the
tachyon matter problem \cite{SenTM}. Another example, in fact a simpler one, to which our results
apply, are deformations generated by $\partial X^{\pm}$. Physically they correspond to turning on
light-like Wilson lines, or in the T-dual picture, where the branes become localized both in space
and time, they describe their separation in the light-like direction. We shall not however expand
on this solution further.

In Section \ref{s_Disc} we propose another type of solution, in what might be called a pseudo
$\bb_0$ gauge. This seems easier to apply to situations with non-trivial self-contractions because
of the absence of certain singularities.

Marginal deformation solutions in open \sft have been studied previously in
\cite{SZmarg,TT,Kluson1,Kluson2,SenEMT,Zeze}, whereas \cite{YZ} initiated their study in closed \sf
theory. In the course of this work we have learned that similar results to ours have been obtained
independently by Kiermaier, Okawa, Rastelli and Zwiebach \cite{KORZ} and should appear in preprint
at the same time as our work.

%%%%%%%%%%%%%%%%%%%%%%%%%%%%%%%%%%%%%%%%%%%%%%%%%%%%%%%%%%%%%%%%%%%%%%%%%%%%%%
\sectiono{Marginal deformations in SFT}
\label{s_marg}
%%%%%%%%%%%%%%%%%%%%%%%%%%%%%%%%%%%%%%%%%%%%%%%%%%%%%%%%%%%%%%%%%%%%%%%%%%%%%%

We shall start by solving the \sft equation of motion
\be\label{eom}
Q_B\Psi + \Psi*\Psi =0
\ee
perturbatively in a parameter $\lambda$. Let us denote $\phi_n$ the coefficient of order
$\lambda^n$, so that
\be
\Psi = \sum_{n=1}^\infty \lambda^n \phi_n.
\ee
At order $\lambda$ we find
\be
Q_B \phi_1 = 0.
\ee
To obtain a non-trivial solution we shall take $\phi_1$ to be a non-trivial element of the
cohomology.\footnote{Note that similar construction can be used to construct the tachyon vacuum in
the wedge state basis \cite{Bernoulli}. Therein one takes $\phi_1=Q_B(B_1^L c_1 \ket{0})$. The
solution appears to be pure gauge for all $|\lambda|<1$, but becomes non-trivial at $\lambda=1$.}
It is well known that for each cohomology class there is a representative of the form
\be
\phi_1 = cJ(0) \ket{0},
\ee
where $J$ is a purely matter operator of conformal dimension one, so that indeed $Q_B\phi_1=0$. The
solution to the equation of motion (\ref{eom}) can be determined recursively order by order using
\be\label{eom-order-n}
Q_B\phi_n = - \left[ \phi_1 \phi_{n-1} + \phi_2 \phi_{n-2}+ \cdots +  \phi_{n-1} \phi_1 \right].
\ee
The right hand side is manifestly $Q_B$ closed, as one can convince themselves by induction, but it
is a-priori not clear whether it is also $Q_B$ exact. It turns out, that for operators which are
exactly marginal in the conformal field theory, the right hand side is always exact.

To invert $Q_B$ on an $Q_B$-exact state we have to fix a gauge. Popular option, which works well in
the level truncation, is the Siegel gauge \cite{SZmarg}; however, for analytic computations it is
more convenient to use the $\bb_0$ gauge introduced in \cite{Bernoulli}, or some of its variants.
In principle one could try using $\bb_0 + \xi \bb_0^\star$ gauge\footnote{The gauge with $\xi=1$
was found very useful in computing scattering amplitudes \cite{Suzuki}.} with $\xi$ different at
each order of $\lambda$, but in this section we shall stick to the simplest $\bb_0$ gauge. We
remind the reader that $\bb_0$ is the zero mode of the $b$-ghost in the cylinder coordinate.

Let us work out in detail the order $\lambda^2$ of the solution. Easy computation using the
formalism of \cite{Bernoulli} gives
\bea
\phi_2 &=& -\frac{\bb_0}{\ll_0} \left( \phi_1*\phi_1 \right)  = -\int_0^1 dr \, r^{\ll_0-1} \bb_0 (\phi_1*\phi_1) \\
\nonumber\\
&=& -\int_2^3 ds \, \Uhat_s \left[\frac{\pi}{4}\left(\tilde c(x) + \tilde c(-x)\right) -
\frac{1}{2} \bbb \tilde c(x) \tilde c(-x) \right] \tilde J(x) \tilde J(-x) \ket{0}
\nonumber\\
&=&
- \frac{1}{2} \int_2^3 ds \, \tilde c \tilde J \ket{0} * \bbb \ket{s-2} * \tilde c \tilde J \ket{0}
\nonumber\\
\label{phi2}
 &=& - \frac{\pi}{2} \int_0^1 dr \,\phi_1*B_1^L \ket{r} * \phi_1,
\eea
where $x$ in the second line stands for $(\pi/4)(s-2)$. The operator $\Uhat_s$, in a notation
borrowed from \cite{Suzuki}, is defined as
\be
\Uhat_s \equiv U_s^* U_s = e^{-\frac{s-2}{2} \Lhat}.
\ee
The operator $U_s$, in turn, is defined as the scaling operator $(2/s)^{\ll_0}$ in the cylinder
coordinate. The star denotes a BPZ conjugate and $\Lhat$ stands for $\ll_0+\ll_0^\star$. For more
properties the reader is referred to \cite{Bernoulli} as well as to older works \cite{RZ, wedge}.
Note, that in the last two expressions, we have formally integrated over wedge states with $r \in
(0,1)$. Such wedge states are ill-defined, have no meaning on their own, but in the present case
they cause no problem. In fact, we use them only for notational convenience to denote a well
defined operation of deleting part of the empty surface from the states $\phi_1$. The real problem
can only arise in the $r \to 0$ limit, where the two insertions of $cJ$ from the two $\phi_1$'s are
approaching each other, squeezing in between a $b$ line integral. For generic matter operators $J$
there would be a singularity.\footnote{The easiest way to deal with the singularity is to introduce
a lower cut-off $\eps$ on the $r$ integral and define $\phi_2$ by the {\it minimal subtraction}.
This amounts to defining $\int_0^1 dr r^{-2} = -1$, which is the right definition for inverting
$\ll_0$ on weight $-1$ state $c_1$. Unfortunately, it turns out that in the $\bb_0$ gauge there is
also an additional $1/r$ singularity associated with a state $Q_B (\lll c_1) \ket{0}$ on which
$Q_B$ cannot be inverted within the $\bb_0$ gauge. The $1/r$ singularity also appears for
non-exactly marginal operators.} For simplicity we shall restrict our discussion in this section to
operators with finite products at coincident points.

\begin{figure}[th]
\begin{center}
\input{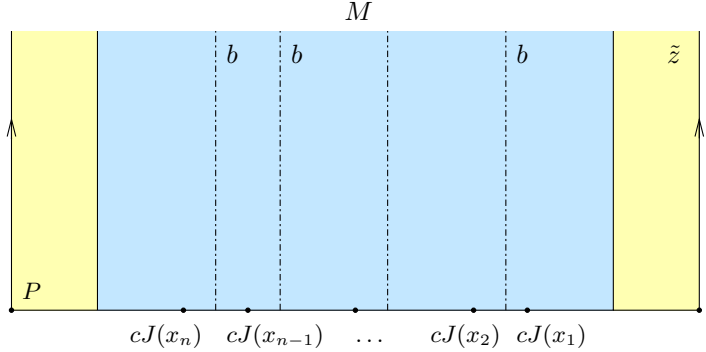}
\caption{\small Representation of the integrand (\ref{sol-guess}) by a cylinder of circumference
$\frac{\pi}{2}\left(2+\sum r_i\right)$ with insertions of BRST nontrivial operator $cJ(z)$ and
$b$-ghost line integral. The cylinder is formed by identifying the lines marked with an arrow. The
BPZ contractions of the integrand are defined as a correlator on this surface with the contracting
vertex operator being inserted at the puncture $P$.} \label{marg_sol}
\end{center}
\end{figure}

Before moving to higher orders in $\lambda$, let us check that indeed
\be
Q_B\phi_2 + \phi_1*\phi_1 =0.
\ee
Acting with $Q_B$ on $\phi_2$ given by (\ref{phi2}) is easy. It annihilates the two factors of
$\phi_1$, and acting on $(\pi/2) B_1^L \ket{r}$ produces $-\partial_r \ket{r}$.  The integral
therefore localizes at the boundary, the $r=1$ contribution gives precisely $-\phi_1*\phi_1$
whereas the $r=0$ contribution vanishes thanks to the two $c$-ghost insertions approaching each
other, again assuming absence of singularity from the matter currents. This is such a simple
mechanism, that it is rather straightforward to guess the form of the $n$-th order term of the
solution
\be\label{sol-guess}
\phi_n = \left(-\frac{\pi}{2}\right)^{n-1} \int_0^1 \prod_{i=1}^{n-1} dr_i \, \phi_1 * B_1^L
\ket{r_1}* \phi_1 * \cdots * B_1^L \ket{r_{n-1}}*\phi_1.
\ee
The proof that this solves (\ref{eom-order-n}) is easy and is left to the reader. Geometric picture
of our solution is given in Fig. \ref{marg_sol}.
The solution $\Psi_\lambda$ can be viewed as a functional on the Hilbert space of the open string,
and as such, it can be represented as a surface with certain punctures. Instead of the conventional
upper-half plane, we use a coordinate where the midpoint is sent to infinity so that the surface
looks as a cylinder of canonical circumference $\pi$. But upon taking star product or acting with
$1/\ll_0$, the natural circumference of the cylinder changes, and in our case, for the solution
(\ref{sol-guess}), it is given by $\frac{\pi}{2}\left(2+\sum r_i\right)$. In addition, we get
insertions of $c J$ located at points
\be\label{defxi}
x_i = \frac{\pi}{4} \left(\sum_{k=1}^{n-1}r_k - 2\sum_{k=1}^{i-1}r_k \right).
\ee

What about the $\bb_0$ gauge condition, does it remain true for our guess (\ref{sol-guess})? Using
the identity
\be
\bb_0 \left(X*Y\right) = -\frac{\pi}{2} (-1)^{{\mathrm gh}(X)} X* B_1^L Y,
\ee
valid for arbitrary $X$ and $Y$ satisfying $\bb_0 X= \bb_0 Y = 0$, we find successively that
$\phi_1 * B_1^L \ket{r_1}$, $\phi_1 *  B_1^L \ket{r_1} *\phi_1, \ldots $ are all in the $\bb_0$
gauge. To see that, one has to use also the identity $B_1^L \ket{r} * B_1^L \phi_1 = 0$.

The tachyon solution was originally found in a similar form \cite{Bernoulli}, but later an elegant
closed form was found by Okawa \cite{Okawa}. In the present case, just by simple inspection, we can
formally sum up the whole series to obtain
\be
\Psi = \frac{\lambda}{1+\frac{\pi}{2}\lambda \int_0^1 \phi_1*B_1^L\ket{r}} * \phi_1.
\ee
Here again, the first factor by itself does not make much sense. However, as it acts on $\phi_1$,
its action is well defined. Actually, it is possible to avoid using the negative wedge states. One
can re-write the formula as
\be
\Psi = \sqrt{\ket{0}}*\frac{\lambda}{1+\lambda \hat\phi*A}*\hat\phi * \sqrt{\ket{0}},
\ee
where $\sqrt{\ket{0}}$ is a star square-root of the vacuum, or in other words the wedge state
$\ket{3/2}$, $\hat\phi=\Uhat_1 c \tilde J(0)\ket{0}$ and finally
\be
A = \frac{\pi}{2} B_1^L \int_1^2 dr \ket{r}
\ee
is the homotopy operator used in \cite{ES} to prove that the cohomology around the tachyon vacuum
is trivial.

Let us now ask what are the interesting properties of the marginal solution. For exactly marginal
deformation one would expect that the energy of the configuration relative to the original brane is
strictly zero. In the time independent setup, the energy is simply given by minus the action. Under
the change of the parameter $\lambda$ the action changes as
\be
\frac{\partial S}{\partial\lambda} = \aver{\frac{\partial\Psi_\lambda}{\partial\lambda}, Q_B
\Psi_\lambda + \Psi_\lambda * \Psi_\lambda} = 0,
\ee
where we used the fact that $\Psi_\lambda$ is a solution of the equations of motion for all values
of $\lambda$. Integrating the equation we find that $S=0$ for all finite values of $\lambda$.

At first sight, there is a little puzzle however. It seems that this proof works not only for
exactly marginal deformations, but for all kinds of one-parameter families of solutions
continuously connected to zero. One may think of a solution generated by a dimension zero operator
\be
A_1^a \, c\partial X^1 \otimes \sigma_a + A_2^b \, c \partial X^2 \otimes \sigma_b
\ee
in a system of two D-branes, where the Chan-Paton factors are given by the Pauli matrices. This
corresponds to turning on a constant non-abelian gauge potential $A=A_1 dx^1 + A_2 dx^2$ along two
directions. As is well known, constant non-abelian fields have nonzero potential energy given by
$\Tr \left[A_\mu, A_\nu\right]^2$; this is true also in \sf theory, as can be shown by integrating
out infinite tower of massive fields \cite{BS}. It turns out, that for such deformations the
recursive procedure for finding the solution breaks down. One has to go back and correct the
initial starting point $\phi_1$ by higher order corrections. Typically what happens is that $e^{i k
X}$ gets changed to $e^{i k (\lambda) X}$ which itself is a nice conformal operator, but its
variation with respect to $\lambda$ is not. This is the point where our formal proof would break
down. The problem does not arise in the fully compact Euclidean case, since there are no operators
with continuous spectrum, and so the obstructions in the recursive procedure are unsurmountable.
From the field theory perspective these obstructions manifest themselves as impossibility to turn
on continuously a flux on a compact manifold.

Another general and interesting question to ask is how the cohomology of the theory changes under
the marginal deformation. Had we worked out in detail, for instance, the solution corresponding to
branes moving apart, we would have to be able to see how does the mass-spectrum of the stretched
strings changes linearly with the brane separation\footnote{This question was touched upon in the
context of string field theory in \cite{BSZ} and \cite{Maccaferri}.}. We do not have the solution
yet, nevertheless, we can address the problem first from a formal viewpoint. Expanding the \sft
around the new vacuum $\Psi_\lambda$, we get the new BRST-like operator
\be
Q_{\lambda} = Q_B + \{\Psi_\lambda, \, \bullet \,  \}_*
\ee
and we want to find its cohomology. Formally, this is actually rather easy to determine. Start with
a solution $\Psi_\lambda$ to the equation $Q_B \Psi + \Psi * \Psi=0$ and perturb it in the
direction of some operator $\varphi$. The variation of the solution solves
\be
Q_\lambda \delta\Psi = Q_B \delta\Psi + \{\Psi_\lambda, \delta\Psi\}_* =0.
\ee
So the cohomology is given by perturbed solutions. These are very easy to construct. Deform the
original theory by an operator $\lambda J(z) + \mu \varphi(z)$, pretending that it is still exactly
marginal operator -- in reality it is not, of course. The solution will be given by the formula
(\ref{sol-guess}), but only its first order term in $\mu$ will be relevant for the new cohomology
representatives. Concretely the solution is
\bea
\ket{O_\varphi} &=& \sum_{n=1}^\infty \left(-\frac{\pi}{2} \lambda\right)^{n-1} \int_0^1
\prod_{i=1}^{n-1} dr_i \, \left[ c\varphi * B_1^L\ket{r_1} * cJ * \cdots *B_1^L \ket{r_{n-1}} * cJ
+ \mbox{\, ($n-1$ terms)} \right]
\nonumber\\
&=& c \varphi \ket{0} -\frac{\pi}{2} \lambda \int_0^1 dr\, \left[ c\varphi * B_1^L\ket{r} * cJ +
 cJ * B_1^L\ket{r} * c\varphi \right] + \cdots,
\eea
where the $n-1$ terms in the first line are obtained by exchanging the position of $\varphi$ with
the remaining $J$'s. It is also possible to rewrite the formula in a closed form
\be\label{Ophi}
\ket{O_\varphi} = \left(1-\Psi_\lambda * B\right) \varphi \left(1-B*\Psi_\lambda\right),
\ee
where
\be
B= \frac{\pi}{2} \int_0^1 B_1^L \ket{r}
\ee
is a formal object, meaningful when sandwiched between two states containing half-strips of size
$\pi/4$ without any insertions on the side adjacent to $B$ . Apart of this purely notational
formality,the solution (\ref{Ophi}) might be jeopardized when the OPE between $J$ and $\phi$ is
singular (which is in fact the typical case). As we have been consistently ignoring these issues,
we will do so once more. We shall postpone them to a future work. It is perhaps interesting that
the straightforward formal proof of (\ref{Ophi}) does not require $\Psi$ to be a marginal
deformation solution. It can be just any solution to the equations of motion. Of course, we do
expect, that in the tachyon vacuum (\ref{Ophi}) will become singular.

%%%%%%%%%%%%%%%%%%%%%%%%%%%%%%%%%%%%%%%%%%%%%%%%%%%%%%%%%%%%%%%%%%%%%%%%%%%%%%
\sectiono{Rolling solutions}
\label{s_roll}
%%%%%%%%%%%%%%%%%%%%%%%%%%%%%%%%%%%%%%%%%%%%%%%%%%%%%%%%%%%%%%%%%%%%%%%%%%%%%%

The most interesting application of the previous results is to the study of rolling tachyons
\cite{SenRolling,SenTM}. Such solutions are generated by a primary field $J=e^{\pm X_0}$ of
dimension one (we are using units in which $\alpha'=1$). For definiteness, we shall take only the
plus sign in the exponent -- so that the tachyon field is in the perturbative vacuum in the far
past. The important property of this vertex operator is that for positive powers its boundary OPE's
are non-singular
\be\label{eXOPE}
:e^{m X^0(x)}: \,  :e^{n X^0(y)}: \; \simeq \; |x-y|^{2 n m} :e^{(m+n) X^0(y)}:.
\ee
To construct the solution we can simply use the results from the previous section, setting
\be
\phi_1 = c_1 e^{X^0} \ket{0}.
\ee
The solution itself is given by (\ref{sol-guess}); in a form more suitable for level truncation
analysis it reads
\be
\Psi = \sum_{n=1}^\infty \lambda^n \left(-\frac{\pi}{2}\right)^{n-1} \! \int \prod_{i=1}^{n-1} dr_i
\, \Uhat_{2+\sum r} \left[-\frac{1}{\pi} \bbb \tilde c(x) \tilde c(-x) + \frac{1}{2} (\tilde c(x)
+\tilde c(-x)) \right] \tilde J(x_1) \ldots \tilde J(x_n) \ket{0},
\ee
where $x\equiv x_1$ and the $x_i$ are given by formula (\ref{defxi}). Now let us extract the
coefficients $c_1 e^{n X_0} \ket{0}$. This will tell us the time dependence of the tachyon field.
This has been previously studied in various approximation schemes in
\cite{SenRolling,SenTM,MZ,Yang,LNT,Lambert,FH1,Moeller,FH2,Bonora,Erler2004,FGN,Coletti}. The
puzzling feature encountered was that the solution, conjectured to be the tachyon matter, was
oscillating with exponentially growing amplitude. The computed pressure was following the same
pattern in stark contrast with Sen rolling tachyon conjectures \cite{SenRolling,SenTM}. Logically,
there seem to be two possible explanations. Either the solution has a finite radius of convergence
in $e^{X^0}$, so that beyond that one has to use proper re-summation formula. An example of such
behavior is $(1+\lambda e^{X_0})^{-1}$, which in fact is quite reminiscent of the results from the
boundary state analysis. Another possible explanation, perhaps the more likely one, is that the
pressure is given by a more complicated formula, containing perhaps some improvement terms that are
not given by the Noether procedure. In that case the oscillations in the tachyon field would not
have any physical meaning.

The coefficient of the state $c_1 e^{n X_0} \ket{0}$ in the rolling solution is given by
\bea\label{coeff}
&& \lambda^n \left(-\frac{\pi}{2}\right)^{n-1} \int \prod_{i=1}^{n-1} dr_i
\left(\frac{2}{2+\sum_{i=1}^{n-1} r_i}\right)^{n^2+n-2} \cos^2 y \left[1-\frac{2y}{\pi}-\frac{1}{\pi}\sin 2y\right]
\nonumber\\
&& \qquad \times \aver{I \circ e^{-n X^0(0)} \tilde J(y_1) \ldots \tilde J(y_n)},
\eea
where $y\equiv y_1$, $\tilde J(y_k) = \cos^{-2}y_k e^{X^0(\tan y_k)}$ and further
\be
y_i^{(n)} = \frac{\pi}{2} - \pi \frac{1+\sum_{k=1}^{i-1} r_k}{2+\sum_{k=1}^{n-1} r_k}.
\ee
The matter correlator can be computed using the OPE (\ref{eXOPE})
\bea
\aver{I \circ e^{-n X^0(0)} \tilde J(y_1) \ldots \tilde J(y_n)} &=& \prod_{i=1}^{n}
\frac{1}{\cos^{2n}y_i} \prod_{1 \le i<j\le n} \sin^2(y_i-y_j)
\nonumber\\
&=& \prod_{i=1}^{n} \frac{1}{\cos^{2}y_i} \prod_{1 \le i<j\le n} \left(\tan y_i - \tan
y_j\right)^2.
\eea
To compute the coefficient (\ref{coeff}) analytically, it is convenient to pass from $r_i$ to the
$y_i$ variables. We were able to compute only the first three coefficients explicitly
\be
\Psi = \left[ \lambda e^{X^0} - \lambda^2 \frac{64}{243 \sqrt{3}} e^{2X^0} + \lambda^3 a_3 e^{3X^0}
+\cdots \right] c_1 \ket{0} + \cdots,
\ee
where $a_3$ is rather complicated expression which depends on polygamma function at special points.
Equivalently, it can be expressed using the Hurwitz zeta functions $\sum_{n=1}^\infty
(n+\alpha)^{-s}$ for $s=2,3,\ldots 9$. This is in fact quite natural, since the conformal dimension
of $e^{3X^0}$ is $9$ and therefore the transcendentality pattern is similar to the one for the
ghost number zero tachyon solution. The value of $\alpha$ runs over the values $0,1/12,2/12,\ldots,
11/12$.

\begin{figure}[t]
\begin{center}
\epsfbox{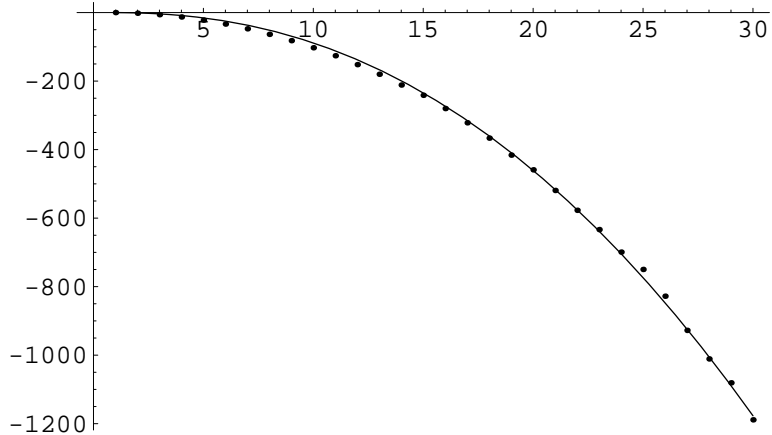} \caption{Absolute value of a logarithm of the coefficients $a_n$
for $ e^{n X^0} c_1 \ket{0}$. One parameter fit works remarkably well.} \label{FigMC}
\end{center}
\end{figure}

Proceeding to higher orders in $\lambda$ analytically seems an impossible task, so we have tried to
obtain number of coefficients numerically by Monte Carlo integration\footnote{Actually we have used
the built-in method QuasiMonteCarlo in Mathematica, so that our approximate values are exactly
reconstructible.}. The first few values we got with $10^7$ points are
\bea
&& a_1 = 1, \quad a_2= -0.152, \quad a_3 = 0.00215, \quad a_4 = -2.62 \, 10^{-6},
\\\nonumber
&& a_5 = 2.79 \, 10^{-10}, \quad a_6 = -2.80 \, 10^{-15}, \quad a_7 = 2.73 \, 10^{-21}, \quad  a_8
= 2.59 \, 10^{-28}.
\eea
With  less accuracy, $10^5$ points, we went up to values of $a_{30}$. The results are plotted in
the graph \ref{FigMC}. They seem to be fitted remarkably well by a one-parameter fit $n^{-0.38
n^2}$.
The behavior seems to be consistent with that of  Moeller and  Zwiebach \cite{MZ} and Fujita and
Hata \cite{FH1,FH2} who also found faster than exponential decay in the coefficients, which means
that the sum over powers of $e^{nX_0}$ has infinite radius of convergence and hence the infinitely
growing oscillations will stay. To be completely honest, we have to point out, that our data are
not entirely conclusive in this respect. Had the integrand been only moderately peeked around some
cube, e.g. of volume $(1/10)^{10}$ for $a_{10}$, there would be virtually no chance of detecting
this region by the Monte Carlo method. The more reliable numerical or even analytic data for lower
order coefficients do not however suggest this scenario. So although we are missing rigorous prove
we have enough evidence to believe that $a_n$ decay faster than exponentially, so that our series
in $e^{X_0}$ has infinite radius of convergence.

%%%%%%%%%%%%%%%%%%%%%%%%%%%%%%%%%%%%%%%%%%%%%%%%%%%%%%%%%%%%%%%%%%%%%%%%%%%%%%
\sectiono{Discussion}
\label{s_Disc}
%%%%%%%%%%%%%%%%%%%%%%%%%%%%%%%%%%%%%%%%%%%%%%%%%%%%%%%%%%%%%%%%%%%%%%%%%%%%%%

We have presented a rather simple solution describing marginal deformation generated by
dimension-one matter primary operator with finite $J(x) J(y)$ as $x \to y$. In order to be able to
study really interesting examples, such as generic rolling tachyon process, or properties of
unstable systems of branes and (anti)branes one has to understand well the case with singular $J(x)
J(y)$. Our preliminary computations show that the otherwise successful $\bb_0$ gauge might not
allow for existence of such solutions. There is one very simple alternative to the construction
presented in section \ref{s_marg}. When taking the $Q_B$ inverse of an $Q_B$ exact object, instead
of demanding that the whole thing be in the $\bb_0$ gauge, we may as well simply demand that the
argument behind $\Uhat_r$ be in the $\bb_0$ gauge\footnote{This trick was invented, as far as we
know, by Ian Ellwood in 2002 in the context of the Siegel gauge. He called this a pseudo-Siegel
gauge.}. For example for $\phi_2$ we find
\be
\tilde\phi_2 = - \frac{1}{2!} \Uhat_3 \int_{-\frac{\pi}{4}}^{\frac{\pi}{4}} dz \, \left(\tilde
c(z)+\tilde c(-z)\right) \tilde J(z) \tilde J(-z) \ket{0}.
\ee
Working out few more terms, it seems that the pattern is
\be
\ket{\Psi} = \sum_{n=1}^\infty (-1)^{n+1} \frac{\lambda^n}{n!} \Uhat_{n+1}
\int\!\!\!\ldots\!\!\!\int_{-\frac{\pi}{4}}^{\frac{\pi}{4}} \prod_{i=1}^n dz_i \, \delta\left(\sum
z_i\right) \theta_{\cal M}(z_i) \sum_{i=1}^n \tilde c(z_i) \prod_{i=1}^n \tilde J(z_i)  \ket{0},
\ee
where $\theta_{\cal M}(z_1,\ldots,z_n)$ is the characteristic function of a domain specified by the
set of inequalities
\be
\left| \sum_{j=1}^k z_{i_j}\right| \le \frac{\pi}{4} k (n-k)
\ee
for $1 \le k \le n-1$. We leave the proof of this proposal for the future.

\section*{Acknowledgments}

I would like to thank Ian Ellwood and Ashoke Sen for useful discussions. I also wish to acknowledge
the hospitality of HRI in Allahabad and the organizers of the Indian Strings Meeting in Puri. This
research has been supported in part by DOE grant DE-FG02-90ER40542.

\end{document}